\documentstyle[prl,aps,12pt,epsf]{revtex}
\def\be{\begin{equation}}
\def\th0{\theta_0}

\def\bea{\begin{eqnarray}}
\def\ee{\end{equation}}
\def\eea{\end{eqnarray}}

\begin{document}
\title{Overlap properties and adsorption transition \\
 of two
Hamiltonian paths }
\author{S. Franz}
\address{Abdus Salam International Center for Theoretical Physics,\\
Strada Costiera 11, P.O. Box 563, 34100 Trieste, Italy}
%\and
\author{T. Garel and H. Orland}
\address{Service de Physique Th\'eorique\\
CE-Saclay, 91191 Gif-sur-Yvette Cedex\\
France}
\date{\today}
\maketitle

\begin{abstract}
We consider a model of two (fully) compact polymer chains, coupled through
an attractive interaction. These compact chains are represented
by Hamiltonian paths (HP), and the coupling favors the existence of common
bonds between the chains. Using a ($n=0$ component) spin representation for
these paths, we show
the existence of a phase transition for strong coupling  (i.e. at low
temperature) towards a ``frozen'' phase where one chain is completely
adsorbed onto the other. 
By performing a Legendre transform, 
we obtain the probability
distribution of overlaps.
The fraction of common bonds between two HP,
i.e. their overlap $q$,  has both lower ($q_m$) and upper ($q_M$)
bounds. This means in particuliar that two HP with overlap greater
than $q_M$ coincide. These results may be of interest in (bio)polymers
and in optimization problems.  
\end{abstract}

\vskip 3mm

\vskip 5mm \noindent\mbox{Submitted for publication to:} \hfill %
\mbox{Saclay, SPhT/98-128}\newline
\noindent \mbox{``EPJ B ''}\newline
\vskip 5mm\noindent \mbox{PACS: 61.41.+e; 87.15.Da; 64.70.Pf} \newpage

\newpage
%\begin{multicols}{2}
%%%%%%%%%%%%%%%%%%%%%%%%%%%

\section{Introduction}

\label{sec: intro} 
The study of two coupled identical systems is a familiar
topic in Statistical Physics, both in equilibrium \cite{Ash_Tel} and
non equilibrium contexts \cite{Der}. It is commonly considered in the physics
of glassy systems \cite{Bla,Par}, where it is a substitute to the time honored
conjugate field in the exploration of phase space. In
this note, we wish to extend these studies to the case of two compact
polymer chains. The coupling between the chains is chosen as an attractive
term which favors the existence of common monomers (bonds); in other
words, we consider the ``adsorption'' of a chain onto the other. 
By a Legendre transform, this model maps onto the overlap probability
distribution of the chains.

For
simplicity, we will consider fully compact chains. This allows us to use a
simple spin representation \cite{de_Ge} to describe the chain properties
through Hamiltonian paths. Beside its interest in homopolymer physics, this
model may yield some insight in the sequence-stucture relationship in
proteins, or in some related optimization problems such as the travelling
salesman. The outline of the paper is as follows: we briefly recall in
section \ref{sec:hampa} the connection between a single polymer chain and
an $n$-component spin system, in the limit $n=0$. The extension to the
problem of two interacting chains (section \ref{sec:twoham}) is then
straightforward, and will be 
solved at a mean field level in section \ref{sec:saddle}. We finally
mention some possible consequences of the model.

\section{The single chain problem}

\label{sec:hampa} The connection between polymer physics and spin systems
may be presented as follows \cite{Sar}. Let us consider, on each site $\vec{%
r}$ of a $d$-dimensional cubic lattice, a spin variable $\vec{S}_{r}$, with $%
n$ components. The normalization is such that $\vec{S}_{r}^{2}=n$. Consider
the sum: 
\begin{equation}
Z(K)=\int \prod_{\vec{r}}d\mu (\vec{S}_{r})\prod_{\left\{ \vec{r},\vec{r}%
^{ }\right\} }\left( 1+K\vec{S}_{r}\vec{S}_{r^{ }}\right) 
\label{O(n)}
\end{equation}
where the variable $K$ denotes the fugacity of a monomer, and the
product runs over neighbouring pairs of sites (i.e. bonds) of the lattice. In
eq  (\ref{O(n)}), $d\mu $ is  the normalized integration measure on
the $(n-1)$ dimensional sphere of radius $\sqrt{n}$. 

\bigskip Due to the normalization of  $\vec{S}_{r}$, it is easy to see
that
\begin{equation}
\label{field}
\int d\mu (\vec{S}_{r})\exp (\vec {H}_{r} \vec {S}_{r}%
)=1+\frac{\vec {H}_{r}^{2}}{2}+O(n)
\end{equation}
so that, by taking derivative with respect to $H_{r}^{\alpha }$ ($\alpha
=1,...,n)$, we obtain:

\[
\int d\mu (\vec{S}_{r})\,1=1;
\;\;\;\;\; 
\int d\mu (\vec{S}_{r})\,(S_{r}^{\alpha })^{2}=1
\]

All higher powers of  $\vec{S}_{r}$ are at least of order $n$.

 Expanding eq (\ref{O(n)}%
) in powers of $K$, and using the previous remark, we see that $Z(K)$ can be
viewed as a sum over all closed loops, each closed loop of length $l$
contributing a weight $n K^{l}$. In the limit $n\rightarrow 0$, only the
single connected loops survive, hence the possibility to represent
self-avoiding walks (SAW) on the lattice by eq (\ref{O(n)}). A very
useful rewriting of  eq (\ref{O(n)}) is
\begin{equation}
Z(K)=\int \prod_{\vec{r}}d\mu (\vec{S}_{r}) \ {\rm e}^{{1 \over 2}\sum_{\left( \vec{r},\vec{r}%
^{ }\right)} K\vec{S}_{r}\Delta _{\vec{r}\vec{r}^{\prime }} \vec{S}_{r^{\prime }}} 
\label{expo}
\end{equation}
where the operator $\Delta _{\vec{r}\vec{r}^{\prime }}$ is a lattice
 $\Delta $  function ($\Delta _{\vec{r}\vec{r}^{\prime }}=1$, if
 $\vec{r}$ and $\vec{r}^{\prime }$ are  neighbouring sites, and
 $0$ otherwise). Note that the sum in the exponential term of
 eq (\ref{expo}) is over the sites of the lattice. Applying
 the familiar Stratonovich-Hubbard  transformation to eq (\ref{expo}),
 we introduce $n=0$ component fields $\vec {\phi}_{r}$ and get
\begin{equation}
Z(K)=\int \prod_{\vec{r}} d\vec {\phi}_r \ {\rm e}^{-{1 \over
2} \sum_{\left( \vec{r},\vec{r}%
^{\prime }\right)}
\vec{\phi}_{r}\Delta _{\vec{r}\vec{r}^{\prime }}^{-1}
\vec{\phi}_{r^{\prime }}}  \int \prod_{\vec r} d\mu
(\vec{S}_{r}) \ {\rm e}^{{\sqrt{K} \sum_{\vec r} \vec{S}_{r} \vec {\phi}_r}}
\label{expo2}
\end{equation}
Using eq (\ref{field}), we get 
\begin{equation}
Z(K)=\int \prod_{\vec{r}} d\vec {\phi}_r \ {\rm e}^{-{1 \over
2} \sum_{\left( \vec{r},\vec{r}%
^{\prime }\right)}
\vec{\phi}_{r}\Delta _{\vec{r}\vec{r}^{\prime }}^{-1}
\vec{\phi}_{r^{\prime }}}  \prod_{\vec r} \left(1+{K \over 2}{\vec {\phi}_r}^{2}\right)
\label{expo3}
\end{equation}
So far we have not specified the spatial extension of the SAW: $Z(K)$
is the grand partition function of the chain, so that the number $M$
of sites of the lattice is not related to the number $N$ of bonds of
the SAW. We now
require that fully compact configurations are the only configurations
present in eq (\ref{expo3}), i.e. $M=N$. This full compactness
requirement amounts to keep the term proportional to $n K^{N}$ in
eq (\ref{expo3}). The coefficient of this term is simply the number
$\cal N$ of self avoiding fully compact configurations (also called
Hamiltonian paths or HP), that is 
\begin{equation}
{\cal N}=K^{-N} \left({Z(K) \over n} \right)_{n=0}={\rm lim}_{n\rightarrow 0}
{1 \over n} \int \prod_{\vec{r}}
d\vec {\phi}_r \ {\rm e}^{-{1 \over 
2} \sum_{\left( \vec{r},\vec{r}%
^{\prime }\right)}
\vec{\phi}_{r}\Delta _{\vec{r}\vec{r}^{\ \prime }}^{-1}
\vec{\phi}_{r^{\prime }}}  \prod_{\vec r} \left({{\vec {\phi}_r}^{2}
\over 2}\right)
\label{expo4}
\end{equation}
Note that ${\cal N}$ is simply the canonical partition function for a
HP of $N$ bonds.
Performing a homogeneous saddle point approximation
on $\vec \phi$ in eq (\ref{expo4}) yields \cite{Or_It_Do}
\begin{equation}
\label{meanfi}
 {\cal N}=\left({z \over e}\right)^{N}
\end{equation}
where $z=2d$ is the coordination number of the lattice. Note that in
the context of Hamiltonian paths, an homogeneous solution implies that
one deals with periodic boundary conditions  \cite{Go_Ka_Ma}. 

Equation (\ref{expo4}) holds for an arbitrarily 
connected graph with adjacence matrix $\Delta_{\vec{r}\vec{r}^{\
\prime}}$, but the saddle  
point evaluation is a priori valid only for large enough $z$.
Indeed, for the fully connected graph ($z=N$), the number of 
HP can be directly estimated  (${\cal N}=N!/(2N)$), and
eq (\ref{meanfi}) reduces to the Stirling formula.   

\section{The coupled chains problem}
\label{sec:twoham}
We now extend the above approach to the case of two interacting
self avoiding chains. By ``interacting chains '', we mean that the
presence of common (or doubly occupied) bonds between the chains is
thermodynamically favored, and we are interested in
counting the number of configurations of two such HP. 

Following eq (\ref{O(n)}), we denote by $K$ the fugacity of a monomer
and consider the sum
\begin{equation}
Z_{2}(K,\lambda)=\int \prod_{\vec{r}}d\mu (\vec{S}_{r})d\mu (\vec{\sigma}_{r})
\prod_{\left\{ \vec{r},\vec{r}%
^{\prime }\right\} }\left( 1+K(\vec{S}_{r}\vec{S}_{r^{\prime
}}+\vec{\sigma}_{r}\vec{\sigma}_{r^{\prime }})+{\lambda} K^2 (\vec{S}_{r}\vec{S}_{r^{\prime }})(\vec{\sigma}_{r}\vec{\sigma}_{r^{\prime }})\right) 
\label{O(2n)}
\end{equation}
where $\vec {\sigma}_{r}$ has the same properties as $\vec {S}_r$
($\vec{\sigma}_{r}^{2}=n$), and the parameter $\lambda$ is a measure
of the interaction between the chains (see below). The contribution of
doubly occupied bonds to eq (\ref{O(2n)}) is proportional to
$\lambda^{N_{12}}$, where ${N_{12}}$ is the number of these
bonds (implying $\lambda>1$).

Following exactly the same steps as in section \ref{sec:hampa}, we
have  
\begin{equation}
Z_{2}(K,\lambda)=\int \prod_{\vec{r}} d\mu (\vec{S}_{r})d\mu (\vec{\sigma}_{r}) \ {\rm e}^{{1 \over 2}\sum_{\left( \vec{r},\vec{r}%
^{\prime }\right)} \Delta _{\vec{r}\vec{r}^{\prime }} G_{2}(\vec
S,\vec {\sigma})}
\label{doub1}
\end{equation}
with 
\begin{equation}
G_{2}(\vec
S,\vec {\sigma})= K(\vec{S}_{r} \vec{S}_{r^{\prime
}}+\vec{\sigma}_{r} \vec{\sigma}_{r^{\prime }})+(\lambda-1)K^{2}
(\vec{S}_{r}\vec{S}_{r^{\prime
}})(\vec{\sigma}_{r}\vec{\sigma}_{r^{\prime }}) 
\label{doub2}
\end{equation}
Performing the Stratonovitch-Hubbard transformation on eqs.(\ref{doub1}) and
(\ref{doub2}) yields   
\begin{equation}
\label{doub3}
Z_{2}(K,\lambda)= \int \prod_{\vec{r}} d\vec {\phi}_r d\vec {\psi}_r
\prod_{\alpha \beta} dq_{\alpha \beta}(r) \ {\rm e}^{-{1 \over
2} \sum_{\left( \vec{r},\vec{r}%
^{\prime }\right)}\Delta_{\vec{r}\vec{r}^{\prime}}^{-1}{\cal
A}_{\vec r \vec r^{\prime}}}
  \int \prod_{\vec r} d\mu (\vec{S}_{r})d\mu (\vec{\sigma}_{r}) \ {\rm
e}^{{\sqrt{K} \sum_{\vec r} {\cal B}_{r}}}
\end{equation}
with
\begin{equation}
\label{doub4}
{\cal A}_{\vec r \vec r^{\prime}}=\vec{\phi}_{r} \vec{\phi}_{r^{\prime
}}+\vec{\psi}_{r} \vec{\psi}_{r^{\prime }}+\sum_{\alpha
\beta}q_{\alpha \beta}(\vec r)q_{\alpha \beta}(\vec r^{\prime})
\end{equation}
and
\begin{equation}
\label{doub5}
{\cal B}_{r}= \vec{S}_{r} \vec {\phi}_{r} +\vec
{\sigma}_{r} \vec {\psi}_{r} +\sqrt{K} \sqrt{\lambda -1} \sum_{\alpha \beta}
q_{\alpha \beta}(r) S_{r}^{\alpha}{\sigma}_{r}^{\beta}
\end{equation}
Note that the previous transformations require two $n$ component
fields $\vec {\phi}_{r}$ and $\vec {\psi}_{r}$, and an $n \times n$ matrix
$q_{\alpha \beta}(r)$ , with $n=0$. One may now perform the
integration over the original spin variables $\vec{S}_{r}$ and
$\vec{\sigma}_{r}$ in eq (\ref{doub3}). We obtain 
\begin{equation}
\label{doub6}
Z_{2}(K,\lambda)= \int \prod_{\vec{r}} d\vec {\phi}_r d\vec {\psi}_r
\prod_{\alpha \beta} dq_{\alpha \beta}(r) \ {\rm e}^{-{1 \over
2} \sum_{\left( \vec{r},\vec{r}%
^{\prime }\right)}\Delta_{\vec{r}\vec{r}^{\prime}}^{-1}{\cal
A}_{\vec r \vec r^{\prime}}} \prod_{\vec r} {\cal C}_{r}
\end{equation}
where
\begin{equation}
\label{doub7}
{\cal C}_{r}=1+{K\over 2}({\vec {\phi}_{r}}^{2}+{\vec {\psi}_{r}}^{2})
+{K^{2} \over 4}{\vec {\phi}_{r}}^{2}{\vec
{\psi}_{r}}^{2}+K^{2}\sqrt{\lambda -1}\sum_{\alpha \beta}
\phi_{r}^{\alpha} q_{\alpha \beta}(r) \psi_{r}^{\beta} +{K^{2} \over
2}(\lambda -1) \sum_{\alpha \beta}q_{\alpha \beta}^{2}(r)
\end{equation}
So far, we have not specified the spatial extension of the
chains. The sum $Z_{2}(K,\lambda)$ is grand canonical with
respect to both singly and doubly occupied bonds. Following the
previous section, we now enforce the full  
compactness constraint for both chains, by keeping the term
proportional to $n^{2} K^{2N}$ in eq (\ref{doub6}).
Defining by ${\cal Z}(\lambda)$, the total number of HP of the two
interacting chains, we have 
\begin{equation}
\label{canon}
{\cal Z}(\lambda)=K^{-2N}\left({Z_{2}(K,\lambda) \over n^{2}}\right)_{n=0}
\end{equation}
Technically, the extraction of the term proportional to $K^{2N}$ in
$Z_{2}(K,\lambda)$, amounts to keep on each site $\vec r$  the terms
proportional to $K^{2}$ in eqs.(\ref{doub6}) and (\ref{doub7}). We may
therefore write
\begin{equation}
\label{estim1}
{\cal Z}(\lambda)={\rm lim}_{n \to 0} {1 \over n^{2}}  \int \prod_{\vec{r}} d\vec {\phi}_r d\vec {\psi}_r
\prod_{\alpha \beta} dq_{\alpha \beta}(r) \ {\rm e}^{-{1 \over
2} \sum_{\left( \vec{r},\vec{r}%
^{\prime }\right)}\Delta_{\vec{r}\vec{r}^{\prime}}^{-1}{\cal
A}_{\vec r \vec r^{\prime}}} \prod_{\vec r} {\cal D}_{r}
\end{equation}
where ${\cal A}_{\vec r \vec r^{\prime}}$ is given in eq (\ref{doub4})
and
\begin{equation}
\label{estim2}
{\cal D}_{r}={{\vec {\phi}_{r}}^{2}{\vec
{\psi}_{r}}^{2} \over 4}+\sqrt{\lambda -1}\sum_{\alpha \beta}
\phi_{r}^{\alpha} q_{\alpha \beta}(r) \psi_{r}^{\beta} +{1 \over
2}(\lambda -1) \sum_{\alpha \beta}q_{\alpha \beta}^{2}(r)
\end{equation}
Note that ${\cal Z}(\lambda)$ is still grand canonical with respect to
the number $N_{12}$ of common bonds between the two
HP, since, as mentionned above, the term with $N_{12}$ common bonds
in ${\cal Z}(\lambda)$ yields a contribution proportional to
$\lambda^{N_{12}}$. 

One may also interpret eq (\ref{canon}) as giving the
partition function of two HP of $N$ bonds, with an
attractive 
interaction energy $\varepsilon$ favoring common bonds, at temperature
$T$, that is
\begin{equation}
\label{canon2}
{Z}(\varepsilon)=\sum_{(HP_1,HP_2)} \ e^{\beta \varepsilon N_{12}}
\end{equation}
 with $\beta={1\over T}$. The identity between eqs.(\ref{canon}) and
(\ref{canon2}), i.e between ${\cal Z}(\lambda)$ and ${Z}(\varepsilon)$,
yields the familiar result $\lambda=e^{\beta \varepsilon}$. Moreover,
equation (\ref{canon2}) allows us to derive bounds for the partition function
${Z}(\varepsilon)$ (or ${\cal Z}(\lambda)$), namely
\begin{equation}
\label{bound}
{\cal N}  e^{ \beta \varepsilon N} \le  {Z}(\varepsilon)
\le {\cal N}^{2} 
\end{equation}
where ${\cal N}$ is given in eq (\ref{meanfi}). From now on, we will
set $\varepsilon=1$.

\section{Saddle point approximation and phase transitions}
\label{sec:saddle}
Since an exact evaluation of eqs (\ref{estim1},\ref{estim2}) seems to
be out of reach, we
will use a saddle point approximation  with respect to the
variables  $\vec {\phi}_{r}$, $\vec {\psi}_{r}$, and $q_{\alpha
\beta}(r)$. This saddle point will be further restricted to be space
independent ($\vec {\phi}_{r}=\vec {\phi}, \vec {\psi}_{r}=\vec
{\psi}, q_{\alpha \beta}(r)= q_{\alpha \beta}$), again implying
periodic boundary conditions.
Setting ${\cal Z}(\lambda)=e^{-N\omega(\lambda)}$, we get
\begin{equation}
\label{saddle}
\omega(\lambda)={\rm Min}_{(\phi^{\alpha},\psi^{\beta},q_{\alpha
\beta})}\left( {1 \over 4d}({\vec {\phi}}^{2}+{\vec
{\psi}}^{2}+\sum_{\alpha \beta}q_{\alpha \beta}^{2})-{\rm Log}{\cal
D}\right) 
\end{equation}
where
\begin{equation}
\label{sadddle2}
{\cal D}={{\vec {\phi}}^{2}{\vec
{\psi}}^{2} \over 4}+\sqrt{\lambda -1}\sum_{\alpha \beta}
\phi^{\alpha} q_{\alpha \beta} \psi^{\beta} +{1 \over
2}(\lambda -1) \sum_{\alpha \beta}q_{\alpha \beta}^{2}
\end{equation}

The saddle point equations are easily solved, by introducing the
quantities $\Phi ={\vec {\phi}}^{2}, \Psi ={\vec {\psi}}^{2},
{\rm R}=\sum_{\alpha \beta} \phi^{\alpha} q_{\alpha \beta}
\psi^{\beta}$, and ${\rm Q}=\sum_{\alpha \beta}q_{\alpha
\beta}^{2}$. Denoting the saddle point values with a subscript
$0$, we obtain
\begin{equation}
\label{sad1}
{\Phi_0 \over 2d}={{1 \over 2} \Phi_0^{2}+ \sqrt{\lambda-1}{\rm
R}_0 \over {\cal D}_0}
\end{equation}
\begin{equation}
\label{sad2}
{{\rm R}_0 \over 2d}={(\lambda-1){\rm R}_0+
\sqrt{\lambda-1}\Phi_0^{2} \over {\cal D}_0} 
\end{equation}
\begin{equation}
\label{sad3}
{{\rm Q}_0 \over 2d}={(\lambda -1){\rm Q}_0 + \sqrt{\lambda-1}{\rm
R}_0 \over {\cal D}_0}
\end{equation}
together with $\Phi_0=\Psi_0$. In eqs. (\ref{sad1}-\ref{sad3}), the
denominator is given by
\begin{equation}
\label{sad4}
{\cal D}_0= {{\Phi}_0^{2} \over 4}+\sqrt{\lambda-1}{\rm R}_0
+{(\lambda -1) \over 2}{\rm Q}_0
\end{equation}
One then gets
\begin{equation}
\label{sad5}
\omega_0(\lambda)={\Phi_0 \over 2d}+{{\rm Q}_0 \over 4d}-{\rm Log}{\cal D}_0
\end{equation}

At this point, one has to resort to a numerical solution of the saddle
point equations. Note that the possible solutions obey simple
equalities or inequalities, such as: $ \Phi_0,{\rm Q}_0 >0, \ \Phi_0+{\rm
Q}_0=4d, \ {\rm R}_0^{2}={\rm Q}_0 \Phi_0^{2}$. 

Our results can be
interpreted in two ways. They first describe the (thermal)
properties of two HP coupled via eq (\ref{canon2}): ${\rm Log}{\cal
Z}(\lambda)=-N \omega(\lambda)$ 
is then, up to a temperature factor, the free energy (with $\lambda=e^{\beta}
$). On the other hand, a Legendre transformation with 
respect to ${\rm Log}\lambda$, gives information on the number
$N_{12}$ of common bonds between the HP, i.e. on their overlap
properties. For convenience, we define $N_{12}=N q$, and
\begin{equation}
\label{legend}
\Theta(q)=\oint {d\lambda \over {2i\pi}\lambda} {\cal Z}(\lambda)
{\lambda}^{-Nq}
\end{equation}
A saddle point evaluation of eq (\ref{legend}) gives
\begin{equation}
\label{prop}
q=-{\partial \omega(\lambda) \over \partial {\rm Log}\lambda}
\end{equation}
Using eqs (\ref{sad1}-\ref{sad5}), we obtain 
\begin{equation}
\label{prop2}
q_0={\lambda \over \lambda -1}{{\rm Q_0} \over 4d}
\end{equation}
We now present our results along both ways.

\subsection{Thermal properties of the coupled Hamiltonian paths}
\label{subsec:therm}
\begin{figure}
\epsfxsize=300pt \centerline{\epsffile{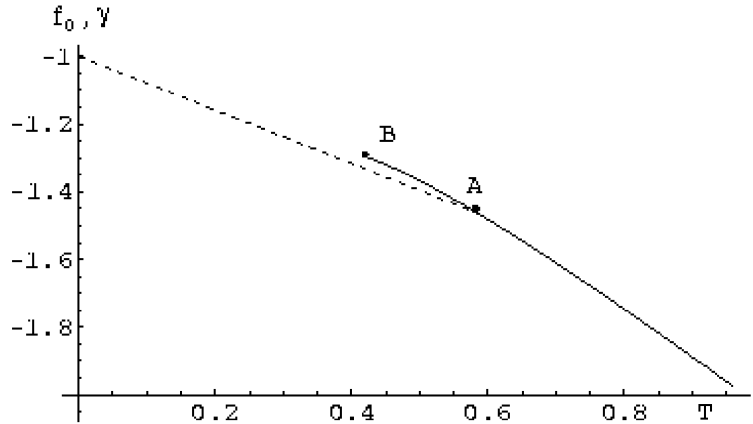}}
\medskip
\caption[0]{\protect\label{fig1} The free energy $f$ as a function 
of temperature. The saddle point result $f_0$ (full line) crosses the
free energy $\gamma$ of the fully adsorbed phase (hatched line) at
point A ($T=T_3$).} 
\end{figure}

Unless otherwise specified, our results are given for $d=3$, and the
main parameter of this section is the temperature $T$ ($T={1 \over
{\rm Log}\lambda}$).  The saddle point equations
(\ref{sad1}-\ref{sad5}) yield the free energy $f_0(T)=T \omega_0(e^{1
\over T})$ as a function of $T$ (see Figure \ref{fig1}).
Solving numerically the above equations, we get the following results

1) $T_{1}=\infty$

In this case, we find a finite fraction of common bonds $q^{(1)}_0={1
\over d}={1 \over 3}$. This value corresponds to a 
random choice of a common bond among the $d=3$ lattice dimensions, and
is the smallest possible value $q_m$ of the overlap.

2) $T_{2}={1 \over {\rm Log}(d+1)} \simeq 0.7213$

One then has $\Phi_0={\rm Q}_0=2d=6$. This point is in some sense a
disorder point \cite{ste} 
where the values of $\Phi_0$ (linked to the entropy of
a single HP) and ${\rm Q}_0$ (linked to the overlap of the two
HP) cross. It corresponds to an overlap  $q^{(2)}_{0}={d+1 \over
2d}={2 \over 3}$.

3) $T_{3} \simeq 0.5846$: the complete adsorption transition.

At this point, the two HP system ``freezes'' into a single HP: one
may also say that one chain gets fully adsorbed onto the other. Of
course, the entropy does not vanish below $T_3$, but becomes equal to
the entropy of a single HP, see eq (\ref{meanfi}). Note that
this solution is not a saddle point solution.
It must nevertheless be taken into account, since its free energy per
monomer which reads
\begin{equation}
\label{free1}
\gamma(T)= -1- T \ {\rm Log}({z \over e})
\end{equation}
is an upper bound (see eq (\ref{bound})). The transition temperature
$T_3$ is thus defined by $\gamma(T_3)=f_0(T_3)$ (see Figure \ref{fig1}). 
The mecanism of this first order freezing transition is analogous to
the one studied in ref.\cite{Ba_Ga_Or}, in the context of polymer
crystallization. Note that $q$ jumps from $q^{(3)}_{0} \simeq 0.7572$
to $q=1$ across the transition. 
From a thermodynamic point of view, we have thus found that the
overlap fraction of two HP has a maximum value $q_M \simeq
0.7572$, beyond which the two HP coincide.

Note that, as the space dimension $d$ increases, the adsorption
temperature $T_3$ decreases. In particular, the transition does not
occur for the fully connected graph ($d \to \infty$).

4) In the present mean field like description, one may continue the high
temperature branch of the free energy $f_0(T)$  beyond the
full adsorption point, implying the existence of a metastable
state (see  the branch AB of Figure \ref{fig1}). In this respect, a
particular temperature 
may be defined, where the entropy $s_0(T)=-{\partial {f}_0(T)
\over \partial T}$ equals that of a single HP:
\begin{equation}
\label{entrop}
s_0(T_4)={\rm Log}({z \over e})
\end{equation}
leading to $T_4 \simeq 0.4350$, and in turn to $q^{(4)}_0 \simeq
0.9627$.
Furthermore, one may argue that this metastable branch is
defined up to the point where the (metastable) overlap parameter $q$
reaches the value one. If this is correct, we find from the saddle
point equation that this occurs for $T=T_5 \simeq 0.3898$, with a positive
entropy $s_0(T_5) \simeq 0.498$. We do not have a clear
understanding of the metastable branch, and more work is
needed on this point.

\subsection{Overlaps and Legendre Transform}
\label{subsec:overlap}
As  previously mentionned, our results can also be interpreted in
terms of the overlap properties of the two HP. A convenient function
to characterize these properties is the overlap probability
distribution  ${\cal P}(q)$ defined as
\begin{equation}
\label{distrib}
{\cal P}(q)= {1 \over {\cal N}^{2}} \sum_{(HP_1,HP_2)}\delta (N_{12}-Nq)
\end{equation}
where $\cal N$ is given in eq (\ref{meanfi}) and $N_{12}$ is the
number of common bonds of the HP. Eqs.(\ref{legend}) and
(\ref{distrib}) imply that 
\begin{equation} 
\label{distrib2}
{\cal P}(q)= {1 \over {\cal N}^{2}} \Theta(q)
\end{equation}

From a strictly thermodynamic point of view, we have the result that
the overlap probability distribution ${\cal P}(q)$ is defined
only for $q_m< q <q_M$, and for $q=1$. The existence of a metastable
branch beyond $q_M$ is not easy to interpret: the saddle point
evaluation (\ref{prop}) of the Legendre 
transform (\ref{legend}) then becomes ill-defined, since $\lambda (q)$
becomes a multivalued function.

For $ q_m < q < q_M$, the relation between $\lambda$ (i.e. temperature) and
$q$ (i.e. overlap) can be inverted through eq (\ref{prop}). There is
thus a one to one correspondance between overlap and temperature 
in the region $1 < \lambda < \lambda_3  = e^{1 \over
T_3} 
\simeq 5.532$.
The thermal properties of the coupled system imply that, for
 $1< \lambda < \lambda_3$, we have  ${\cal
 P}(q)=\delta(q_0-q)$, where $q_0$ is given by
 eq (\ref{prop2}). 
On the other hand, 
for $\lambda > \lambda_3$, we have  ${\cal P}(q)=\delta(1-q)$.

\begin{figure}
\epsfxsize=300pt \centerline{\epsffile{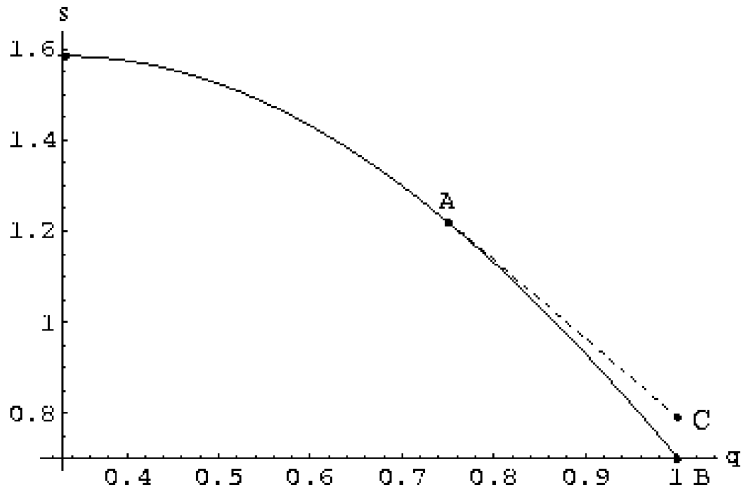}}
\medskip
\caption[0]{\protect\label{fig2} The entropy $s(q)$ as a function 
of the overlap $q$. The full line denotes the saddle point result
$s_0(q)$; the hatched line is the result of the Maxwell 
construction with  the fully adsorbed state. Note
that AB corresponds to the metastable branch of Figure
\ref{fig1}.}   
\end{figure}

Another quantity of interest is the entropy, considered as a function
of the overlap $q$. It is given by
\begin{equation}
s(q) ={1 \over N} \ {\rm Log} \Theta(q)
\end{equation}
Performing the saddle point evaluation on $\lambda$ in eq
(\ref{legend}), we get 
\begin{equation}
\label{entrop2}
s_0(q)= -\omega_0(\lambda)-q {\rm Log} {\lambda}
\end{equation}
where $\lambda=\lambda(q)$ is given by eq (\ref{prop2}). The phase
transition for $q=q_M$ can be interpreted (see Figure \ref{fig2}) as a
Maxwell construction since the results  of the previous section can be
rewritten as 
\begin{equation}
\label{maxwell}
s_0(q_M)-s_0(1)= (1-q_M) \ {\rm Log} \lambda_3 
\end{equation}

Beyond the homogeneous
saddle point approximation, our results in the region $q_M<q<1$ are
coherent with a phase coexistence picture between a fraction $x$ of
the phase $q=q_M$ and a fraction $1-x$ of the  fully adsorbed phase
$q=1$. Such a phase coexistence will give rise to an effective overlap
$q_M<q_{eff}= x q_M+(1-x)<1$. 

\section{Conclusion}
\label{sec:conclu}
We have considered the overlap and/or thermal properties of two
coupled Hamiltonian paths (HP), in a homogenous saddle point
approximation. We have found (for $d=3$) a phase transition at finite coupling,
between an entropy dominated phase and a completely adsorbed phase,
where the two HP have overlap $q=1$. This phase transition disappears
when $d \to \infty$.
Our results may be relevant in a proteic context (NMR, structure
alignement). If the number of constraints exceeds a certain threshold,
the existence of $q_M$ suggests that a single spatial structure may
survive. For longer polymers, the existence of a first order
transition, raises many questions (metastability, glass
transition,..). 
Finally, we remark that the coupled chains system of this paper
undergoes a phase transition at a finite value of the coupling
constant $\lambda$, whereas in spin glasses, the two replica system undergoes
a (spin glass) transition when the coupling constant vanishes
\cite{Bla,Par}. 

S. Franz thanks the Service de Physique Th\'eorique, Saclay,  for its kind
hospitality during the elaboration of this work.

%\end{multicols}

\newpage

\begin{references}
\begin{center}
{\bf REFERENCES}
\end{center}
\bibitem{Ash_Tel} J. Ashkin and E. Teller, Phys. Rev., {\bf 64}, 178,
(1943). For a review, see R.J. Baxter, {\it Exactly Solved Models In
Statistical Mechanics}, Academic Press, New York, 1982.
\bibitem{Der} B. Derrida, Phys. Repts., {\bf 189}, 207, (1989).
\bibitem{Bla} A. Blandin, J. Phys. (France), {\bf 39}, C6, 1499,
(1978).
\bibitem{Par} G. Parisi, Phys. Rev. Lett., {\bf 50}, 1946, (1983).
\bibitem{de_Ge} P.G. de Gennes, Phys. Lett. A, {\bf 38}, 339, (1972).
\bibitem{Sar} G. Sarma in {\it The Ill-Condensed Matter}, Les Houches
1979, R. Balian {\it et al.} eds, North Holland, page 537.
\bibitem{Or_It_Do}  H. Orland, C. Itzykson and C. De Dominicis, J.
Phys. (France) {\bf 46}, L353, (1985).
\bibitem{Go_Ka_Ma}
M. Gordon, P. Kapadia and A. Malakis, J. Phys. A, {\bf 9}, 751 (1976).
\bibitem{ste} J. Stephenson, Can. J. Phys., {\bf 48}, 1724 (1970).
\bibitem{Ba_Ga_Or}  J. Bascle, T. Garel and H. Orland, J.Phys. A, {\bf 25},
L1323, (1992).
\end{references}
\end{document}